# Quadrics for structuring space-time wavepackets

Pierre Béjot [*] and Bertrand Kibler

*Laboratoire Interdisciplinaire Carnot de Bourgogne, UMR6303 CNRS-UBFC, 21000 Dijon, France*

Space-time light structuring has emerged as a very powerful tool for controlling the propagation dynamics of pulsed beam. The ability to manipulate and generate space-time distributions of light has been remarkably enhanced in past few years, letting envision applications across the entire spectrum of optics. Space-time optical wavepackets manipulated up to now are usually two-dimensional objects (one space dimension and time) whose mode-resolved spectra lie in a conical section. Using simple symmetry and invariance principles, we show that such wavepackets are particular cases of more general three-dimensional structures whose space-time frequencies lie on quadric surfaces. Our proposed framework allows here classifying space-time wavepackets localized in all dimensions, in any group-velocity dispersion regime, both in bulk and waveguides. Particular emphasis is placed on orbital angular momentum-carrying space-time wavepackets. This unprecedented theoretical approach opens the way for versatile synthesizing of space-time optics.

**Introduction.** Symmetry and invariant principles play a fundamental role in modern physics [1-2]. Major examples from classical mechanics are the total energy, and linear and angular momenta obeying respective conservation law associated with continuous symmetries (time translation, space translation and rotation). The group theory is the prominent mathematical approach for studying symmetries in various branches ranging from solid state physics to quantum physics. It allows to analyze symmetries and their implications in a unified way for distinct physical systems, including solving or simplifying various problems, such as obtaining analytic solutions of differential equations (e.g. Lie's method). Application of symmetries in optical physics can be easily found in geometrical optics and optical aberrations [3-4], as well as the derivation of non-diffracting solutions in free space [5-7]. For the latter, the symmetry group associated with Helmholtz equation and the coordinate systems in which variables separate give rise to various solution families of invariant or accelerating waves and beams such as plane waves, Bessel and Mathieu beams, and Weber waves [8-10]. Orbital angular momentum (OAM)-carrying beams (i.e., vortex beams) are other possible monochromatic solutions (not always diffraction-free) associated to screw axis symmetry [11]. More generally, structured light or particle beams are nowadays indispensable tools in numerous fields of science and technology [12-18], and a new step forward in the advanced control of wave propagation would facilitate the emergence of new functionalities.

While all the above examples have been deeply studied in the monochromatic regime, strong interest is currently shown for space-time invariant wavepackets. Among them, one can cite space-time light sheets [19], needle pulses [20], X- and O-waves [22-24]. Structuring the space-time characteristics of a field turns out to be a very versatile way to engineer the pulse propagation dynamics, allowing, for instance, to control the group velocity [19,25] or group velocity dispersion [26] of the shaped wavepackets. Not only used for sculpting "at will" the propagation dynamics, the framework of space-time wavepackets has provided a deeper understanding of the mechanisms responsible for supercontinua generation during the nonlinear propagation of ultrashort pulses [22,27,28]. In a geometrical point of view, all the aforementioned studied space-time structures are two-dimensional and share the particularity to own a frequency-resolved modal decomposition that lies on a conic section. Since extremely fast progresses are currently made for generating more and more complex three-dimensional light structures [29], it then seems particularly important at this point to provide an exhaustive classification of three-dimensional invariant space-time wavepackets.

In the present work, we provide a general approach for structuring three-dimensional light ultrashort pulses in both bulk and structured dispersive media. We show that, in two different space coordinates systems (Cartesian and cylindrical), space-time wavepackets that are invariant under uniform rectilinear-motion and spatial rotation operations can be built from the coherent superposition of modes that lie on quadric surfaces in the frequency-resolved modal space, whose nature depends on the considered dispersion regime. The geometrical study of the considered quadrics then allows constructing invariant space-time wavepackets propagating with arbitrarily chosen group velocity and rotation, every two-dimensional space-time wavepackets studied so far simply becoming particular cases of the proposed framework.

***Propagation equation in the modal representation.*** The partial differential equation driving the linear propagation in bulk media of an electric field $E$ is given by:

$$[\partial_z^2 + \Delta_\perp + k^2(\omega)]\tilde{E}(\vec{r}_\perp, z, \omega) = 0, \quad (1)$$

where $k(\omega) = \frac{n(\omega)\omega}{c}$, $c$ is the light velocity in vacuum, $n(\omega)$ is the refractive index at the angular frequency $\omega$, $\Delta_\perp$ is the transverse Laplacian operator, $\vec{r}_\perp$ is the transverse position vector, and $\tilde{E}$ is the Fourier transform of $E$ with respect to the time coordinate. The very first step for studying z-invariant space-time wavepackets is to express the eigenvectors (called hereafter "modes") of the operator $\square_\perp : \Delta_\perp + \frac{n^2(\omega)\omega^2}{c^2}$. In the Cartesian (resp. cylindrical) case, they simply correspond to monochromatic plane waves $e^{ik_y y} e^{ik_x x} e^{-i\omega t}$ [resp. orbital angular momentum Bessel monochromatic waves $J_l(k_\perp r) e^{il\theta} e^{-i\omega t}$] with eigenvalues $K_z^2 = \frac{n^2(\omega)\omega^2}{c^2} - k_\perp^2$, where $k_\perp^2 = k_x^2 + k_y^2$. The evolution of a field $E$ along the $z$ direction is given by:

$$\partial_z \bar{E} = iK_z \bar{E}, \quad (2)$$

where $\bar{E}$ is the decomposition coefficients of the field in the basis composed of the eigenvector of $\square_\perp$. Any electric field can be then represented in the modal basis $(k_x, k_y, \omega)$ in Cartesian coordinates and $(k_r, l, \omega)$ in cylindrical coordinates.

***Galilean-transformed (uniform rectilinear motion) invariant wavepackets.*** The first example of invariant space-time wavepackets are those propagating without any deformation in a frame propagating at their group velocity $v_g$. In this case, the transformation linking the field at a given propagation distance $z$ with the initial field is simply $t' \rightarrow t - z/v_g$. Finding the spatiotemporal wavepackets invariant under uniform rectilinear motion (and invariant under phase transformation), amounts to find the fields that belong to the kernel of the differential operator $\partial_z + K_1 \partial_t - iK_0$, where $1/K_1$ is the group velocity of the built wavepacket and $K_0$ is its propagation constant. It immediately follows that the decomposition of rectilinear motion-invariant wavepackets only embeds eigenvectors whose eigenvalues respect

$$K_z = K_0 + K_1 \omega. \quad (3)$$

One can now study how the mode-resolved spectrum of such invariant fields are structured in bulk media. For simplicity, we will consider only second-order dispersion around a given arbitrarily chosen frequency $\omega_0$ and will use the paraxial approximation. Note that such approximations do not change the underlying physics but it allows dealing with more tractable analytical formula. Under these assumptions, Eq. 3 reads

$$\frac{k_\perp^2}{2k_0} - \frac{k_2}{2}\left(\Omega + \frac{k_1 - K_1}{k_2}\right)^2 - C = 0, \quad (4)$$

where $\Omega = \omega - \omega_0$, $k_0 = k(\Omega = 0)$, $k_1 = \partial_\omega k(\Omega = 0)$, and $k_2 = \partial_\omega^2 k(\Omega = 0)$, and $C = (k_0 - K_0) - \frac{(k_1 - K_1)^2}{2k_2}$. The above algebraic expression is a quadric of dimension 2, whose surface depends on the considered dispersion regime (normal or anomalous), the sign of $C$, and the considered coordinates system. Table I summarizes the distinct regimes and associated quadric surfaces of rectilinear motion-invariant wavepackets.

- *Case 1: normal dispersion* ($k_2 > 0$).
Figure 1(a-b) shows the corresponding surfaces for negative and positive values of $C$ for the Cartesian case [i.e., in the $(k_x, k_y, \omega)$ space] namely hyperboloids of one or two sheets. When this term is null, then we find an elliptic cone as shown in Fig. 1(c). As shown in subplots (a-c), the term $C$ mainly governs the existence of phase-matched frequencies around the central wavelength of the wavepacket. This effect is also clearly visible on the limiting case of space-time wave sheets [see conic sections defined as the intersection of the quadric with the plane $k_y = 0$ in Fig. 1(e-g)]. Such programmable correlations between the spatiotemporal degrees of freedom have recently been studied in detail in Refs. [19,25,30,31]. Such waves take the forms of different hyperbola in the $(k_x, \omega)$ and $(k_x, K_z - k_0 - k_1 \omega)$ planes with baseband or passband features, while the relation between $K_z$ and $\omega$ is simply linear, whose slope corresponds to the chosen group velocity. In those subplots, in the vicinity of conic sections, we also point out the local curvature of phase matching induced by dispersion, in strong contrast to standard light cones studied in Ref. [19,25,30,31] restricted to free-space propagation of narrowband wavepackets.

| Quadric surface | Dispersion regime | $C$ | Coordinates system |
|---|---|---|---|
| Hyperboloid one sheet | Normal | $> 0$ | Cartesian |
| Hyperboloid two sheets | Normal | $< 0$ | Cartesian |
| Elliptic cone | Normal | 0 | Cartesian |
| Ellipsoid | Anomalous | $> 0$ | Cartesian |
| Hyperbolic cylinder | Normal |  | Cylindrical |
| Elliptic cylinder | Anomalous | $> 0$ | Cylindrical |
| Ø | Anomalous | $\leq 0$ | Cartesian/Cylindrical |

Table I: Rectilinear motion-invariant wavepackets in bulk media.

In the case of cylindrical coordinates, quadric surfaces [in the $(k_\perp, l, \omega)$ space] naturally embed the typical patterns of X-wave (or conical wave) solutions previously studied in bulk [23-24] or guiding [22,32] media, obtained as the intersection of the quadric surfaces and the plane defined by $l = 0$.

- *Case 2: anomalous dispersion* ($k_2 < 0$).

In that dispersion regime, there is only one possibility to fulfill the phase matching, namely when $C > 0$. The associated quadric is an ellipsoid in Cartesian coordinates, as shown in Fig. 1(d) or an elliptic cylinder in cylindrical coordinates. In the latter case, one retrieves the typical signature of O-waves (light bullets) previously studied in bulk media [23-24]. Again, the limiting case of space-time wave sheet can be obtained for $k_y = 0$ in Fig. 1(h). This wave takes the form of distinct ellipses in the $(k_x, \omega)$ and $(k_x, K_z - k_0 - k_1\omega)$ planes, while the linear relation between $K_z$ and $\omega$ is again driven by the chosen group velocity.

***Galilean-transformed (uniform motion and rotation) invariant wavepackets.*** One can push further the classification of invariant space-time wavepackets by noticing that rectilinear motion-invariant fields form a subgroup of the group of fields which are not only invariant under uniform motion but also invariant at a given rotation around the $z$ axis. These latter, called spatiotemporal helicon wavepackets [28], belong to the kernel of the space-time screw axis symmetry differential operator $\Pi = \partial_z + K_1\partial_t + K_l L_z - iK_0$, where $K_l$ is an arbitrarily chosen constant and $L_z = x\partial_y - y\partial_x = \partial_\theta$ is the z-component of the angular momentum operator. In the following, the study of helicon space-time wavepackets is treated in the more general framework of cylindrically symmetric waveguides, i.e., in media whose refractive index $n$ may exhibit a radial dependence. The bulk case can then be seen as the limiting case of waveguide of infinite dimension.

For simplicity, we will restrict our study to a scalar approach by considering the weak guidance approximation. The partial differential equation driving the linear propagation of an electric field $E$ is then given in cylindrical coordinates by:

$$\left[\partial_z^2 + \partial_r^2 + \frac{1}{r}\partial_r + \frac{1}{r^2}\partial_\theta^2 + \frac{n^2(r,\omega)\omega^2}{c^2}\right]\tilde{E}(r,\theta,z,\omega) = 0, \quad (5)$$

where $n(r, \omega)$ is the radial-dependent refractive index at $\omega$. By using the separation of transverse and longitudinal variables, the linear evolution of the field along the propagation axis is given in the modal basis by [33]:

$$\partial_z \bar{E} = iK_z(l,p,\omega)\bar{E}, \quad (6)$$

where $K_z(l, p, \omega)$ is the propagation constant of the OAM mode $(l, p)$ at the frequency $\omega$. Recall that $l$ and $p$ refer here to azimuthal and radial indices, where $l(0, \pm 1, \pm 2, \pm 3, ...)$ is the topological charge, related to the phase front of OAM modes. The modes form an orthonormal basis so that any electric field $E$ can be expressed as

$$E(r, \theta, t) = \int \sum_{l,p} \bar{E}(l, p, \omega) A_{l,p}(r, \omega) e^{i(l\theta - \omega t)} d\omega, \quad (7)$$

where $A_{l,p}(r, \omega)$ is the transverse envelope of the mode $(l, p, \omega)$ and $\bar{E}(l, p, \omega)$ are the electric field coordinates in the modal basis. Looking for fields that belong to the kernel of $\Pi$ immediately implies that the decomposition of such wavepackets only embeds a family of eigenvectors whose eigenvalues respect

$$K_z(l, p, \omega_{lp}) = K_0 + K_1 \omega_{lp} + K_l l. \quad (8)$$

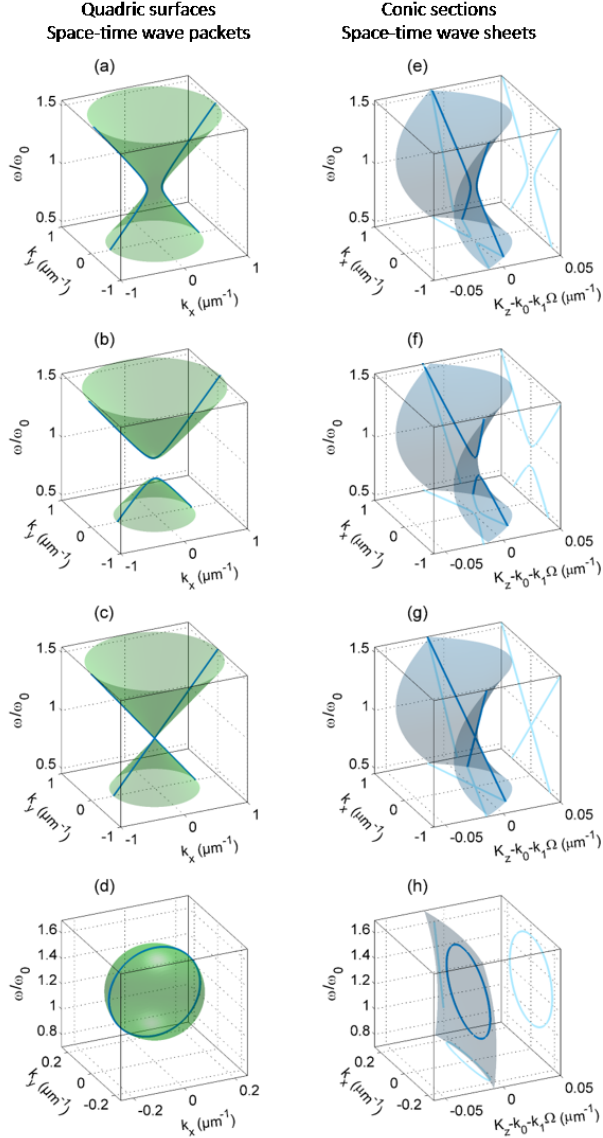

FIG. 1. Quadric surfaces (from Eq. 4) governing phase-matching of (uniform rectilinear motion) invariant space-time wavepacket components in bulk fused silica: (a-c) for normal dispersion with central wavelength fixed at 800 nm, (d) for anomalous dispersion regime with central wavelength fixed at 1600 nm. (a) Hyperboloid of one sheet obtained with $(k_0 - K_0) = 13.85$ cm$^{-1}$ and $(k_1 - K_1) = 7.5$ ps/m. (b) Hyperboloid of two sheets obtained with $(k_0 - K_0) = 13.85$ cm$^{-1}$ and $(k_1 - K_1) = 12.5$ ps/m. (c) Elliptic cone obtained with $(k_0 - K_0) = 13.85$ cm$^{-1}$ and $(k_1 - K_1) = 10$ ps/m. (d) Ellipsoid obtained with $(k_0 - K_0) = 13.85$ cm$^{-1}$ and $(k_1 - K_1) = 10$ ps/m. In subplots (a-d), blue lines indicate the limiting case of space-time light sheets when $k_y = 0$. Corresponding conic sections are reported in subplots (e-h) with their projections onto the distinct planes. Local curvatures induced by dispersion are mapped in the vicinity of conic sections.

Accordingly, any electric field built from a given family is a diffraction- and dispersion-free space-time wavepacket propagating at the group velocity $1/K_1$ whose intensity continuously rotates around the propagation axis with a spatial period $2\pi/K_l$. Before studying how the mode-resolved spectrum of such invariant fields is structured in the modal space $(l, p, \omega)$, one has to emphasize that, generally speaking, the propagation constant in a waveguide (nonlinearly) depends on the topological charge $l$. Hence, still considering only second-order dispersion around a given arbitrarily chosen frequency $\omega_0$, the propagation constant of a waveguide can be well approximated as

$$K_z \simeq k_0 + k_1 \Omega + \frac{k_2}{2} \Omega^2 + \Gamma(l, p, \Omega), \quad (9)$$

with $\Gamma$ is a three-dimensional second-order polynomial:

$$\Gamma(l, p, \Omega) = \gamma_{0,0,0} + \gamma_{1,0,0}|l| + \gamma_{0,1,0} p + \gamma_{0,0,1} \Omega \\ + \gamma_{2,0,0} l^2 + \gamma_{0,0,2} \Omega^2 + \gamma_{0,2,0} p^2 \\ + \gamma_{1,1,0}|l|p + \gamma_{1,0,1}|l|\Omega + \gamma_{0,1,1} p\Omega$$

where the coefficients $\gamma_{i,j,k}$ depend on the geometry of the considered waveguide. For the case of homogeneous bulk media, all coefficients vanish except $\gamma_{0,2,0}$ with $p$ being replaced by $k_r$. Note that the propagation constant usually does not depend on the sign of $l$. This will have some consequences in the geometric nature of the surfaces describing spatiotemporal helicon wavepackets. Inserting Eq. 9 in Eq. 8, it follows that spatiotemporal helicon wavepackets are contained in surfaces whose algebraic expression are

$$\frac{(k_2 + 2\gamma_{0,0,2})}{2}\left(\Omega + \frac{k_1 - K_1 + \gamma_{0,0,1}}{k_2 + 2\gamma_{0,0,2}}\right)^2 + \Delta(l, p, \Omega) - K_l l + C = 0, \quad (10)$$

with $C = (k_0 - K_0 + \gamma_{0,0,0}) - \frac{(k_1 - K_1 + \gamma_{0,0,1})^2}{2(k_2 + 2\gamma_{0,0,2})}$,

and $\Delta(l, p, \Omega) = \gamma_{1,0,0}|l| + \gamma_{0,1,0} p + \gamma_{2,0,0} l^2 + \gamma_{0,2,0} p^2 + \gamma_{1,1,0}|l|p + \gamma_{1,0,1}|l|\Omega + \gamma_{0,1,1} p\Omega$.

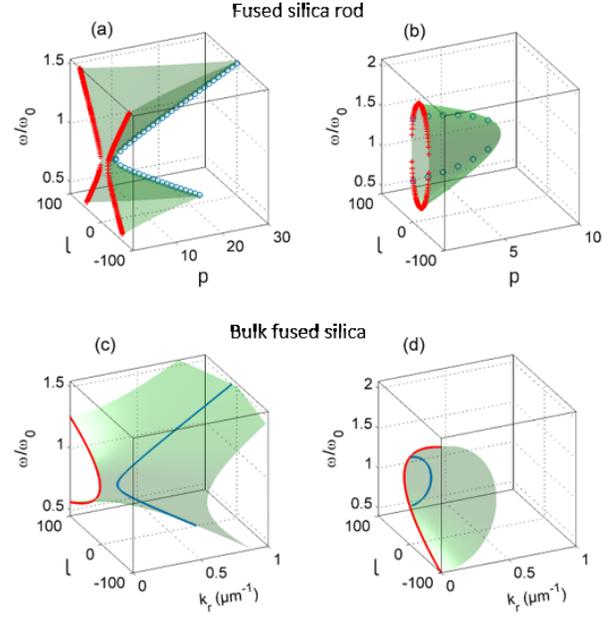

FIG. 2. Examples of quadric surfaces (from Eq. 10) governing phasematching of space-time helicon wavepackets in a fused silica rod with radius $R = 100$ µm (a-b) and bulk fused silica (c-d). Subplots (a: Two hyperboloids of one sheet, c: hyperbolic paraboloid) correspond to the normal dispersion regime with central wavelength fixed at 800 nm, and $(k_0 - K_0) = 13.85$ cm$^{-1}$, $(k_1 - K_1) = 7.5$ ps/m, $2\pi/K_l = 5$ cm. Subplots (b,d) correspond to the anomalous dispersion regime with central wavelength fixed at 1600 nm. $(k_0 - K_0) = 13.85$ cm$^{-1}$, $(k_1 - K_1) = 7.5$ ps/m, and $2\pi/K_l = 5$ cm in subplot (b: Two hyperboloids of two sheets. $(k_0 - K_0) = 13.85$ cm$^{-1}$, $(k_1 - K_1) = 12.5$ ps/m, $2\pi/K_l = 1$ cm in subplot (d: elliptic paraboloid). Blue lines/circles indicate conic sections regarding the limiting cases of non-rotating space-time wavepackets (without superposed OAM) when $l = 0$. Red lines/crosses indicate the simple cases when $k_r = 0$ or $p = 1$.

First, since $\Delta(l, p, \Omega)$ depends on $|l|$, it is straightforward to notice that the surface embedding spatiotemporal wavepackets will be composed of two different quadrics (one for $l > 0$ and another for $l < 0$). However, it appears that the rank and signature (and consequently the nature) of the two involved quadrics is in fact the same. The only difference between the two will be the position of their respective principal axis. Then, it will be sufficient to study the nature of the quadric for positive $l$ only. To this purpose, one has to study the associated matrix $M_Q$:

$$M_Q = \begin{pmatrix} \gamma_{2,0,0} & \gamma_{1,1,0}/2 & \gamma_{1,0,1}/2 \\ \gamma_{1,1,0}/2 & \gamma_{0,2,0} & \gamma_{0,1,1}/2 \\ \gamma_{1,0,1}/2 & \gamma_{0,1,1}/2 & \frac{k_2 + 2\gamma_{0,0,2}}{2} \end{pmatrix}. \quad (11)$$

The relative signs of the eigenvalues $(\lambda_1, \lambda_2, \lambda_3)$ of $M_Q$ completely classify the nature of the quadric surface in the frequency-resolved modal space.

As an example, we investigate below space-time wave structures in a simple and analytical case at the frontier between bulk media and fibers, namely a dispersive medium of finite transversal dimension (radius $R$). In the paraxial approximation, the propagation constants can be approximated by $K_z(l,p,\omega) \simeq k_0 + k_1\Omega + \frac{k_2}{2}\Omega^2 - \frac{\alpha_{lp}^2}{2k_0 R^2}$, where $\alpha_{lp}$ is the $p^{th}$ root of $l^{th}$ Bessel function of first kind $J_l$. In this example, we assume that the usual wavevector $k(\omega)$ in the last term is simplified into $k_0$. Using the formalism described above, with a second-order polynomial fit of $\alpha_{lp}^2$, one can approximatively find the following eigenvalues of the matrix $M_{Q,rod}$:

$$\lambda_1 \simeq -\frac{24.3}{4k_0 R^2}, \lambda_2 \simeq \frac{1.69}{4k_0 R^2}, \lambda_3 = \frac{k_2}{2}. \quad (12)$$

Accordingly, the quadric surface associated to spatiotemporal helicon wavepackets in finite bulk media has a rank of 3. Moreover, since $\lambda_1\lambda_2 < 0$, the quadric surfaces for $l > 0$ and $l < 0$ are necessarily either one- or two-sheet hyperboloids or elliptic cones. Corresponding typical examples are shown in Fig. 2(a-b) in both dispersion regimes. Note that for a clearer illustration, surfaces are plotted even if only discrete values can be found for parameters $(l,p,\omega)$. Discrete values are then shown for limited cases corresponding to conic sections when $l = 0$ or $p = 1$. In the limiting case of infinite bulk media, the rank of the associated matrix decreases to 2 since the propagation constant does not depend on $l$ anymore. Consequently, in this case, the quadrics involved for structuring spatiotemporal helicon wavepackets will be necessarily an elliptic/hyperbolic paraboloid, depending on the respective value of $k_2$, see Fig. 2(c-d). Note that for $K_l = 0$, we retrieve an elliptic/hyperbolic cylinder.

***Numerical construction of wavepackets.*** Next, we present some examples illustrating our general approach of synthesizing rotating space-time optical wavepackets through quadrical phasematching of their spatiotemporal components (Eq. 10). In particular, we construct such space-time wave structures in the simple case of a fused silica rod with radius $R = 100$ μm, both in normal and anomalous dispersion regimes. As a result, for each couple $(l,p)$, one must find the roots of the quadric, thus corresponding to frequencies $\Omega_{lp}$ that depict the family of phase-matched modes $(l,p,\Omega_{lp})$. Figure 3(a) shows the overall three-dimensional pattern of phasematching obtained in the normal dispersion regime, when considering a central wavelength $\lambda_0 = 2\pi c/\omega_0 = 800$ nm, and the arbitrary choice of the constants $K_1 = k_1$ and $K_0 = k_0$ (i.e., propagation constants of the wavepacket are those of $\omega_0$). The rotation period is here fixed equal to $2\pi/|K_l| = 1$ cm. The linear construction of a helicon wavepacket then results from the superposition of the calculated frequencies $\Omega_{lp}$. Figure 3(b) presents an example corresponding to a particular case of selected modes with various radial and angular indices ($0 \leq l \leq 8$, and $1 \leq p \leq 8$), and with equal spectral amplitudes and phases (see circles in Fig. 3a). The successive subplots display the isosurface of the spatiotemporal intensity at half-maximum at distinct propagation distances. The wavepacket looks like a corkscrew in space-time coordinates with extreme localization since all frequencies are in phase. It clearly reveals the spiraling trajectory of the intensity pattern while maintaining its initial distribution over propagation. This helicon wavepacket does not disperse and simply rotates during its propagation around the z- and t-axes because of its inherent invariant nature. The direction of rotation is driven by the sign of $K_l$. Note that complex engineered 3D-patterns can be obtained by a careful selection of multiple higher-order modes at the phase-matched discrete frequencies.

Another example of quadrical phasematching obtained in the anomalous dispersion regime is shown in Fig. 3(c), when considering a central wavelength $\lambda_0 = 2\pi c/\omega_0 = 1800$ nm, and the arbitrary choice of the constants $K_1 = k_1$ and $K_0 = k_0$. The rotation period is now fixed equal to $2\pi/|K_l| = 1.25$ cm. Figure 3(d) presents an example of constructed helicon wavepacket obtained from the superposition of modes with the following radial and angular indices: $0 \leq l \leq 22$, and $1 \leq p \leq 2$, and with equal spectral amplitudes and phases (see circles in Fig. 3c). Again we retrieve the typical invariant spiraling trajectory of the intensity pattern. However, we here notice the more complex spatiotemporal pattern made of three consecutive high-intensity sub-pulses with distinct spatial arrangements.

***Discussion and conclusion.*** The concept of space-time wavepackets has emerged as a very promising and powerful tool for controlling the propagation dynamics of pulsed electromagnetic fields. By finely tailoring the modal distribution of the spectral components embedded space-time wavepackets, one can control in large extent its trajectory and velocity, but also the way they spread both in time and space, and this, whatever the dispersion regime. This degree of control let envision extremely exciting applications and new all-optical functionalities. Here, by invoking symmetry arguments, it was shown that all space-time invariant wavepackets reported to date are particular cases of more general three-dimensional objects whose modal spectral distributions lie on quadrics surfaces. With the extremely rapid progresses made for manipulating space-time wavepackets [34], we hope this work will inspire a wealth of applications. Beyond this simple but far-reaching concept of structured light for controlling the linear pulse propagation dynamics, note that the study of quadrics waves, the general concept embedding conical waves, will also help in understanding the physics underlying the mechanism of supercontinua generation.

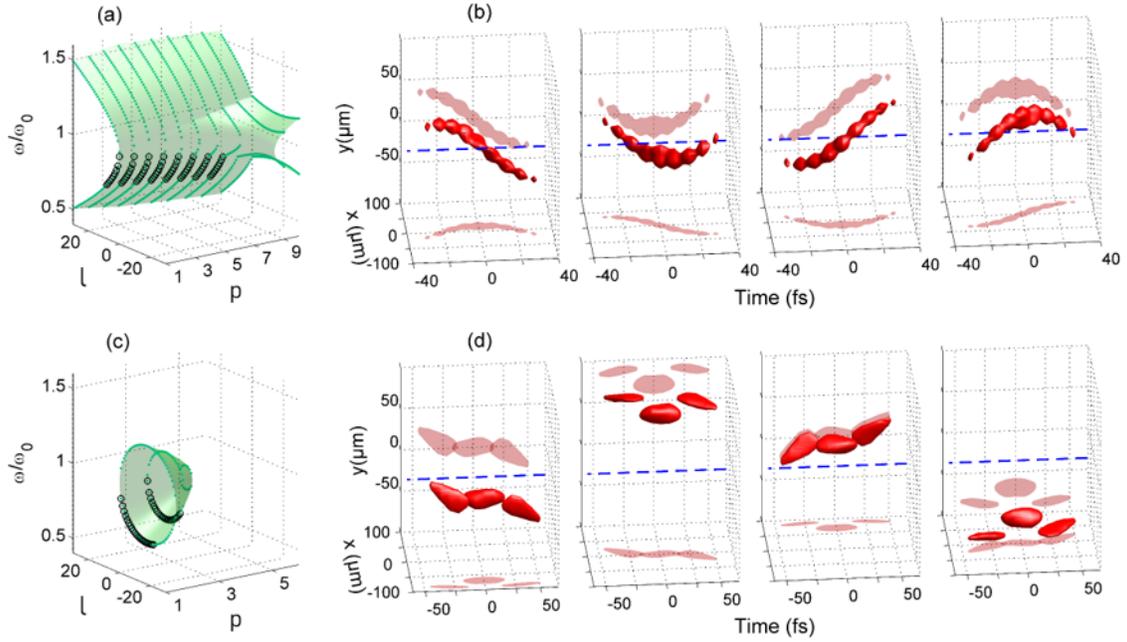

FIG. 3. Examples of helicon wavepackets constructed from quadrical phasematching of spatiotemporal components (from Eq. 10) in a fused silica rod with radius $R = 100$ µm. (a,c) Quadric surfaces obtained both in normal and anomalous dispersion regimes. Green dots indicate discretized conic sections for each $p$ coordinate. Black circles correspond to the selected modes used for the linear construction of helicon wavepackets depicted in (b,d) through iso-surfaces of their spatiotemporal intensity pattern at half-maximum. The successive subplots obtained every $\pi/(2|K_l|)$ confirm the invariant nature of helicon wavepackets over propagation. The dashed blue line indicates the origin ($x = 0, y = 0$). Projections on planes (shadow plots) are also provided for a clear observation of rotation.

For instance, a recent numerical work [28] has already reported that the supercontinuum generated by a helicon pulse is naturally structured in a quadric wave that respects the underlying symmetry imposed by the nonlinear propagation of the pump. Finally, the formalism of quadrical phasematching is not limited to electromagnetic fields and could find applications in various branches of wave physics [35-39], such as acoustics, polaritonics, plasma, gravity, and Bose condensate physics, in order to explore the dynamics of three-dimensional spiraling wavepackets with topological properties.

***Conflict of interest.*** The authors declare that they have no conflict of interest.

***Acknowledgments.*** We acknowledge financial support of the French "Investissements d'Avenir" program (PIA2/ ISITE-BFC, contract ANR-15-IDEX-03; EIPHI Graduate School, contract ANR-17-EURE-0002).

**References**

[1] E. P. Wigner, "Symmetry and conservation laws," Phys. Today **17**, 34 (1964).

[2] D. J. Gross, "The role of symmetry in fundamental physics," Proc. Natl. Acad. Sci. **10**, 14256 (1996).

[3] S. A. Khan, "Group theory in geometrical optics," Jpn. J. Appl. Phys. **17**, 161 (1978).

[4] A. J. Dragt, "Lie algebraic theory of geometrical optics and optical aberrations," J. Opt. Soc. Am. **72**, 372 (1982).

[5] C. P. Boyer, E. G. Kalnins, and W. Miller, Jr, "Symmetry and separation of variables for the Helmholtz and Laplace equations," Nagoya Math. J. **60**, 35 (1976).

[6] A. Wünsche, "Generalized Gaussian beam solutions of paraxial optics and their connection to a hidden symmetry," J. Opt. Soc. Am. A **6**, 1320 (1989).

[7] B. M. Rodríguez-Lara, R. El-Ganainy, and J. Guerrero, "Symmetry in optics and photonics: a group theory approach," Sci. Bull. **63**, 244 (2018).

[8] J. Durnin, "Exact solutions for nondiffracting beams. I. The scalar theory," J. Opt. Soc. Am. A **4**, 651 (1987).

[9] J. C. Gutiérrez-Vega, M. D. Iturbe-Castillo, and S. Chávez-Cerda, "Alternative formulation for invariant optical fields: Mathieu beams," Opt. Lett. **25**, 1493 (2000).

[10] M. A. Bandres and B. M. Rodríguez-Lara, "Nondiffracting accelerating waves: Weber waves and parabolic momentum," New J. Phys. **15**, 013054 (2013).

[11] Y. Shen, X. Wang, Z. Xie, C. Min, X. Fu, Q. Liu, M. Gong, and X. Yuan, "Optical vortices 30 years on: OAM

manipulation from topological charge to multiple singularities," Light. Sci. Appl. **8**, 90 (2019).

[12] H. Rubinsztein-Dunlop et al., "Roadmap on structured light," J. Opt. **19**, 013001 (2017).

[13] A. Forbes, M. de Oliveira, and M. R. Dennis, "Structured light," Nat. Photon. **15**, 253 (2021).

[14] M. Piccardo et al., "Roadmap on multimode light shaping," J. Opt. **24,** 013001 (2022).

[15] S. M. Lloyd, M. Babiker, G. Thirunavukkarasu, and J. Yuan, "Electron vortices: Beams with orbital angular momentum," Rev. Mod. Phys. **89**, 035004 (2017).

[16] M. Urrutia and R. L. Stenzel, "Helicon waves in uniform plasmas. IV. Bessel beams, Gendrin beams, and helicons", Phys. Plasmas **23**, 052112 (2016).

[17] K. Y. Bliokh and F. Nori, "Spin and orbital angular momenta of acoustic beams," Phys. Rev. B **99**, 174310 (2019).

[18] G. M. Vanacore, G. Berruto, I. Madan, E. Pomarico, P. Biagioni, R. J. Lamb, D. McGrouther, O. Reinhardt, I. Kaminer, B. Barwick, H. Larocque, V. Grillo, E. Karimi, F. J. García de Abajo, and F. Carbone, "Ultrafast generation and control of an electron vortex beam via chiral plasmonic near fields," Nat. Materials **18**, 573 (2019).

[19] H. E. Kondakci and A. F. Abouraddy, "Diffraction-free space-time light sheets," Nat. Photon. **11**, 733 (2017).

[20] K. J. Parker and M. A. Alonso, "Longitudinal iso-phase condition and needle pulses," Opt. Express **24**, 28669 (2016).

[21] C. Vetter, T. Eichelkraut, M. Ornigotti, and A. Szameit, "Generalized radially self-accelerating Helicon beams," Phys. Rev. Lett. **113**, 183901 (2014).

[22] B. Kibler and P. Béjot, "Discretized conical waves in multimode optical fibers," Phys. Rev. Lett. **126**, 023902 (2021).

[23] D. Faccio, A. Couairon, and P. Di Trapani, *Conical Waves, Filaments and Nonlinear Filamentation Optics* (Aracne, Roma, 2007).

[24] H. E. Hernandez-Figueroa, M. Zamboni-Rached, and E. Recami, *Localized Waves* (Wiley, Hoboken, 2008).

[25] H. E. Kondakci and A. F. Abouraddy, "Optical space-time wave packets having arbitrary group velocities in free space," Nat. Commun. **10**, 929 (2019).

[26] M.Yessenov, L.A. Hall, and A.F. Abouraddy, "Engineering the Optical Vacuum: Arbitrary Magnitude, Sign, and Order of Dispersion in Free Space Using Space−Time Wave Packets", ACS Phot. **8**, 2274-2284 (2021).

[27] M. Kolesik, E. M. Wright, and J. V. Moloney, "Interpretation of the spectrally resolved far field of femtosecond pulses propagating in bulk nonlinear dispersive media," Opt. Express **13**, 10729-10741 (2005).

[28] P. Béjot and B. Kibler, "Spatiotemporal Helicon Wavepackets", ACS Photonics **8**, 2345–2354 (2021).

[29] C. Guo, M. Xiao, M. Orenstein, and S. Fan, "Structured 3D linear space–time light bullets by nonlocal nanophotonics", Light Sci. Appl. **10**, 160 (2021).

[30] B. Bhaduri, M. Yessenov, and A. F. Abouraddy, "Anomalous refraction of optical spacetime wave packets," Nat. Photon. **14**, 416 (2020).

[31] M. Yessenov, B. Bhaduri, H. E. Kondakci, and A. F. Abouraddy, "Classification of propagation-invariant space-time wave packets in free space: Theory and experiments," Phys. Rev. A **99**, 023856 (2019).

[32] K. Tarnowski, S. Majchrowska, P. Béjot, and Bertrand Kibler, "Numerical modelings of ultrashort pulse propagation and conical emission in multimode optical fibers", J. Opt. Soc. Am. B **38**, 732-742 (2021)

[33] P. Béjot, "Multimodal unidirectional pulse propagation equation", Phys. Rev. E **99,** 032217 (2019).

[34] M. Yessenov, J. Free, Z. Chen, E. G. Johnson, M. P. J. Lavery, M. A. Alonso, and A. F. Abouraddy, arXiv:2111.03095v1

[35] T. Brunet, J-L. Thomas, R. Marchiano, and F. Coulouvrat, "Experimental observation of azimuthal shock waves on nonlinear acoustical vortices," New J. Phys. **11**, 013002 (2009).

[36] L. Dominici, D. Colas, A. Gianfrate, A. Rahmani, V. Ardizzone, D. Ballarini, M. De Giorgi, G. Gigli, F. P. Laussy, D. Sanvitto, and N. Voronova, "Full-Bloch beams and ultrafast Rabi-rotating vortices," Phys. Rev. Res. **3**, 013007 (2021).

[37] J. Vieira, J. T. Mendonça, and F. Quéré, "Optical control of the topology of laser-plasma accelerators," Phys. Rev. Lett. **121**, 054801 (2018).

[38] J. Pierce, J. Webste, H. Larocque, E. Karimi, B. McMorran, and A. Forbes, "Coiling free electron matter waves," New J. Phys. **21**, 043018 (2019).

[39] F. A. Asenjo and S. A. Hojman, "Nondiffracting gravitational waves", Eur. Phys. J. C **81**, 81-98 (2021).